\documentstyle[aps,prl]{revtex}

\newcommand{\BE}{\begin{equation}}
\newcommand{\EE}{\end{equation}}
\newcommand{\BA}{\begin{eqnarray}}
\newcommand{\EA}{\end{eqnarray}}

\begin{document}
\draft
%\tightenlines

\twocolumn[\hsize\textwidth\columnwidth\hsize\csname@twocolumnfalse\endcsname
\title{Propagation inhibition and wave localization in
a 2D random liquid medium}
\author{Yu-Yu Chen and Zhen Ye}
\address{Wave Phenomena Laboratory, Department of Physics, National Central University, Chung-li,
Taiwan 320, ROC}

\date{September 8, 2001}

\maketitle

\begin{abstract}

Acoustic propagation and scattering in water containing many
parallel air-filled cylinders is studied. Two situations are
considered and compared: (1) wave propagating through the array of
cylinders, imitating a traditional experimental setup, and (2)
wave transmitted from a source located inside the ensemble. We
show that waves can be blocked from propagation by disorders in
the first scenario, but the inhibition does not necessarily imply
wave localization. Furthermore, the results reveal the phenomenon
of wave localization in a range of frequencies.

\end{abstract}
]

%\twocolumn
%\narrowtext

When propagating through media containing many scatterers, waves
will be repeatedly scattered by each scatterer, forming a multiple
scattering process\cite{Ishimaru}. Multiple scattering of waves is
responsible for many fascinating phenomena such as random
laser\cite{Laser}, electronic transport in impure solids\cite{Im},
and photonic or acoustic bandgaps\cite{Band,Sanchez,Kush}. Under
proper conditions, multiple scattering leads to the unusual
phenomenon of wave localization. That is, waves in a randomly
scattering medium are trapped in space and will remain confined
around the initial transmitting site until dissipated.

Over the past twenty years, tremendous efforts have been devoted
to the investigation of the localization phenomenon of classical
waves in random media (e.~g.
Ref.~\cite{Kirk1,Genack,Microwave,Ad,weak1}). Observation of
classical wave localization is a difficult task, partially because
suitable systems are hard to find and partially because
observation is often complicated by such effects as absorption and
attenuation. In most previous experimental studies, the apparatus
is set up in such a way that waves are transmitted at one end of a
scattering ensemble, then the scattered waves are recorded either
on the other end to measure the transmission or at the
transmitting site to measure the reflection from the sample. The
results are subsequently compared with the previous theory to
infer possible localization effects. In this way, observations of
wave localization effects have been reported, for example, for
microwaves\cite{Microwave}, acoustic waves\cite{weak1}, and
arguably for light\cite{Light,Argue1,Argue2}, showing propagation
inhibition and an exponential decay in wave transmission along the
propagation path, the indications for wave localization, and
tending to support the prevailing view that waves are localized in
two dimensional (2D) systems with any amount of disorders for all
frequencies\cite{gang4}.

The purpose of this Letter is twofold. First, we would like to
point out that traditional experimental methods have uncertainties
in discerning localization effects, as the observation can be
obscured by effects like reflection and deflection. These effects
attenuate waves, resulting in a similar decay in transmission and
thus making the data interpretation ambiguous. We show that while
wave localization does lead to an inhibition in wave propagation,
but the propagation inhibition does not necessarily imply wave
localization. In other words, it is necessary to differentiate the
situation that waves are blocked from transmission from the
situation that waves can be actually localized in the medium.
Second, we show that in the system we study, although waves are
not localized for all frequencies, wave localization is evident in
a range of frequencies and when there is a sufficient density of
random scatterers.

The model in this Letter is acoustic propagation in water
containing many parallel air-filled cylinders. Different from the
common approach that derives approximately a diffusion equation
for the ensemble-averaged energy, our method is to solve the wave
propagation from the fundamental wave equation, without resort to
approximations. The model has been studied previously for the
coherent behavior of acoustic propagation\cite{EY} and the
acoustic complete bandgaps\cite{Ye3}.
%The advantage of this model
%not only lies in its simplicity, but also in the fact that it is
%exactly computable.

Consider $N$ straight cylinders located  at $\vec{r}_i$ with $i=1,
2, \cdots, N$ to form either a random or regular array. An
acoustic line source transmitting monochromatic waves is placed at
$\vec{r}_s$. The scattered wave from each cylinder is a response
to the total incident wave composed of the direct wave from the
source and the multiply scattered waves from other cylinders. The
final wave reaching a receiver located at $\vec{r}_r$ is the sum
of the direct wave from the source and the scattered waves from
all the cylinders. Such a scattering problem can be solved {\it
exactly}, following Twersky\cite{Twersky}. While the details are
in \cite{Yep}, the essential procedures are summarized below.

The scattered wave from the $j$-th cylinder can be written as \BE
\label{eqps1} p_s(\vec{r}, \vec{r}_j) = \sum_{n=-\infty}^{\infty}
\mbox{i}\pi A_n^j H_n^{(1)}(k|\vec{r} -
\vec{r}_j|)e^{\mbox{i}n\phi_{\vec{r}- \vec{r}_j}}, \EE where $k$
is the wavenumber of the medium, $H_n^{(1)}$ is the $n$-th order
Hankel function of the first kind, and $\phi_{\vec{r}- \vec{r}_j}$
is the azimuthal angle of the vector $\vec{r}- \vec{r}_i$ relative
to the positive $x$-axis. The total incident wave around the
$i$-th cylinder ($i=1,2,\cdots, N; i\neq j$) is the summation of
the direct incident wave from the source and the scattered waves
from all other scatterers, and can be expressed as \BA
\label{eqpin2} p_{in}^i(\vec{r}) &=&
p_0(\vec{r}) + \sum_{j\neq i}^N p_s(\vec{r},\vec{r}_j) \nonumber\\
& =& \sum_{n = -\infty}^\infty B_n^i J_n(k|\vec{r} -
\vec{r_i}|)e^{\mbox{i}n\phi_{\vec{r} - \vec{r_i}}}. \EA

To solve for $A_{n}^i$ and $B_n^i$, we express the scattered wave
$p_s(\vec{r}, \vec{r_j})$, for each $j \neq i$, in terms of the
modes with respect to the $i$-th scatterer by the addition theorem
for Bessel functions\cite{addition}. The resulting formula for the
scattered wave $p_s(\vec{r}, \vec{r_j})$ is \BE \label{eqps2}
p_s(\vec{r}, \vec{r_j}) = \sum_{n=-\infty}^\infty C_n^{j, i}
J_n(k|\vec{r} - \vec{r_i}|)e^{\mbox{i}\phi_{\vec{r} - \vec{r_i}}},
\EE with \BE C_n^{j,i} = \sum_{l=-\infty}^\infty \mbox{i}\pi A_l^j
H_{l-n}^{(1)}(k|\vec{r_i} - \vec{r_j}|)
e^{\mbox{i}(l-n)\phi_{\vec{r_i} - \vec{r_j}}}. \label{Cn}\EE The
direct incident wave around the location of the $i$-th cylinder
can be expressed in a Bessel function expansion with respect to
coordinates centered at $\vec{r}_i$ \BE \label{eqp0exp}
p_0(\vec{r}) = i\pi H_0^{(1)}(kr) = \sum_{l=-\infty}^{\infty}
S_l^i J_l(k|\vec{r} - \vec{r_i}|) e^{\mbox{i}l\phi_{\vec{r} -
\vec{r_i}}}, \EE with the known coefficients
$S_l^i$\cite{addition}.

%$ S_l^i = \mbox{i}\pi H_{-l}^{(1)}(k|\vec{r_i} -\vec{r}_s|) e^{-\mbox{i}l\phi_{\vec{r_i}}}$.

Using equations (\ref{eqpin2}), (\ref{eqps2}) and (\ref{eqp0exp}),
we have \BE\label{eqmatrix1} B_n^i = S_n^i + \sum_{j=1,j\neq i}^N
C_n^{j, i}.\EE The boundary conditions state that the pressure and
the normal velocity be continuous across the interface between a
scatterer and the surrounding medium, leading to \BE \label{BA}
B_n^i = \mbox{i}\pi\Gamma_n^i A_n^i, \EE where $\Gamma_n^i$ are
the transfer matrices relating the acoustic properties of the
scatterers and the surrounding medium, and have been given by
Eq.~(21) in Ref.~\cite{Yep}.
%\BE \Gamma_n^i = \frac{H_n^{(1)}(k a^i) J_n'(k
%a^i/h^i) - g^i h^i H_n^{(1)\prime}(k a^i) J_n(k a^i/h^i)} {g^i h^i
%J_n'(k a^i) J_n(k a^i/h^i) - J_n(k a^i) J_n'(k a^i/h^i)}. \EE Here
%the primes refer to taking derivative, $a^i$ is the radius of the
%$i$-th cylinder, $g^i = \rho_1^i/\rho$ is the density ratio, and
%$h^i = k/k_1^i = c_1^i/c$ is the sound speed ratio for the $i$-th
%cylinder.

The coefficients $A_n^{i}$ and $B_n^j$ can be inverted from
Eqs.~(\ref{Cn}), (\ref{eqmatrix1}), and (\ref{BA}). Once $A_n^{i}$
are determined, the transmitted wave at any spatial point is given
by \BE \label{final} p(\vec{r}) = p_0(\vec{r}) + \sum_{i=1}^N
\sum_{n=-\infty}^{\infty} \mbox{i}\pi A_n^i H_n^{(1)}(k|\vec{r} -
\vec{r}_i|)e^{\mbox{i}n\phi_{\vec{r}- \vec{r}_i}}. \EE The
acoustic intensity is represented by the squared module of the
transmitted wave.

\begin{figure}
\epsfxsize=2.85in\epsffile{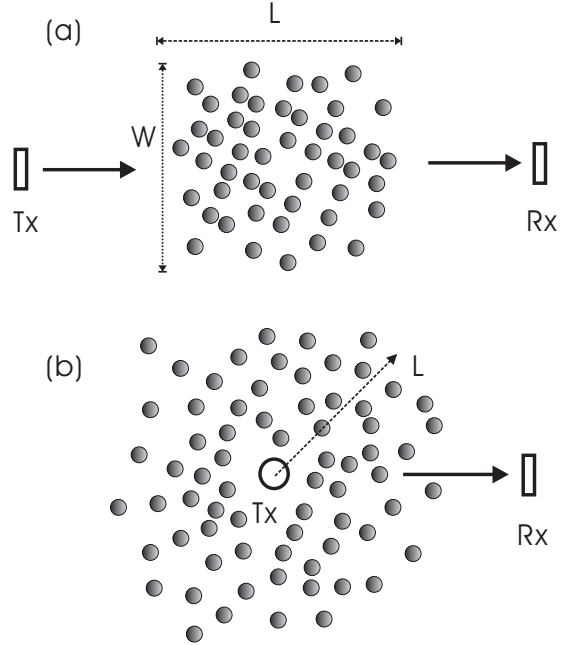} \caption{Conceptual layout:
(a) acoustic propagation through a cloud of cylinders; (b)
acoustic transmission from a line source located inside a cylinder
cloud.} \label{fig1}
\end{figure}

In the following computation, we assume $N$ uniform air-cylinders
of radius $a$. The fraction of area occupied by the cylinders per
unit area is $\beta$. The average distance between nearest
neighbors is therefore $d = (\pi/\beta)^{1/2}a$, which is also the
lattice constant for the corresponding square lattice array. Two
situations are considered: (1) wave propagating through the array
of cylinders, labelled hereafter as the `Outside' situation that
imitates the traditional experimental setup, and (2) wave
transmitted from a source located inside the ensemble, labelled
hereafter as the `Inside' situation. Both cases are illustrated by
Fig.~\ref{fig1}. For the `Outside' case, all cylinders are
randomly placed within a rectangular area with length $L$ and
width $W$. The transmitter and receiver are located at some
distance from the two opposite sides of the scattering area. For
the `Inside' situation, all cylinders are placed within a circle
of radius $L$ with the transmitting source located at the center
and the receiver located outside the scattering cloud. In the
computation, the acoustic intensity is normalized in such a way
that its value equals unity when there are no scatterers present;
thus the uninteresting geometrical spreading effect is naturally
eliminated. We scale all lengths by the parameter $d$, and the
frequency in terms of non-dimensional $ka$; in this way, the
computation becomes non-dimensional.

\input epsf.tex

A set of numerical computations has been performed for various
area fractions $\beta$, numbers $N$, and dimensionless frequency
$ka$. The major controlling parameter is $\beta$. The transmitter
and receiver are placed at a distance of $2d$ from the sample; in
fact, we found that as long as we keep the symmetry the results
remain qualitatively unchanged as the positions of transmitter and
receiver vary. Though it is the line source that is used in
computation, the features hold even when a beamed cylindrical
plane wave is used.

\input epsf.tex
\begin{figure}
\epsfxsize=3in\epsffile{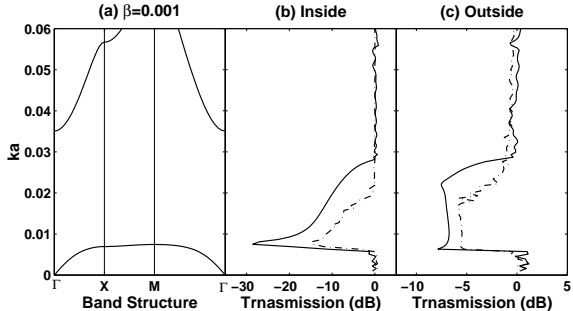} \caption{The middle and right
panels, referring to the `Inside' and `Outside' cases
respectively, show the normalized acoustic transmission versus
frequency in terms of non-dimensional $ka$. Here the comparison is
made between the results from the corresponding square arrays
(solid lines) and from the complete random array of cylinders
(dotted lines). Left panel: The band structures computed by the
plane wave expansion method.} \label{fig2}
\end{figure}

Fig.~\ref{fig2} presents typical results for the transmitted
intensity as a function of $ka$ for the two situations in
Fig.~\ref{fig1}, with $\beta = 10^{-3}$ and $N = 200$. For
reference, we also plot the band structure, following
\cite{Kush,Ye3}, and the transmission for each corresponding
square lattice with the same $\beta$. Here is shown that for both
situations, there is a significant transmission reduction regime
from $ka = 0.007$ to about $0.022$, which is roughly coincident
with the complete bandgap shown by Fig.~\ref{fig2}(a), and within
this regime, the transmission is less inhibited compared to that
from the corresponding square lattice arrays. For the `Inside'
case, the reduction regime for the random scattering is identified
as the localization range. This reduction regime will be widened
as $\beta$ increases, but will disappear when $\beta$ drops below
about $10^{-5}$. Comparing the results from the random arrays and
that from the corresponding square lattice arrays, a significant
difference is apparent: outside the severe reduction regime, the
transmission is reduced by the randomization in the `Outside'
case, for example at $ka=0.05$, but stays nearly unchanged in the
`Inside' scenario. More explicitly, the randomness tends to block
the wave propagation outside the gap regime. Traditionally, such
propagation inhibition caused by randomness has been regarded as
the indication of wave localization. In what follows, we show that
waves are actually not localized at these frequencies.

\begin{figure}
\epsfxsize=3in\epsffile{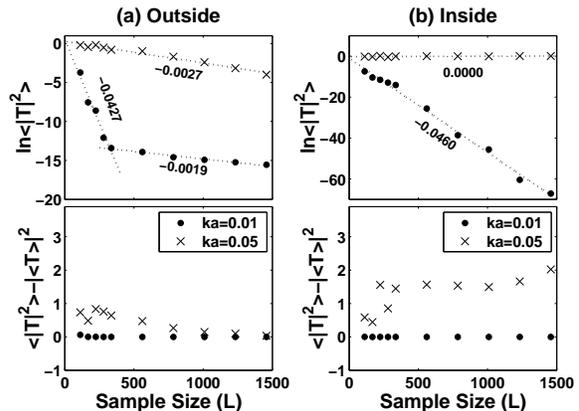} \caption{Normalized acoustic
transmission and its fluctuation versus the sample size for two
frequencies: the left and right panels refer to the `Inside' and
`Outside' cases respectively. The estimated slops for the
transmission are indicated in the figure. Here $T\equiv p/p_0$.}
\label{fig3}
\end{figure}

We consider two frequencies as an example: $ka = 0.01$ and $0.05$.
Fig.~\ref{fig3} presents the results for the random ensemble
averaged transmission and its fluctuation as a function of the
sample size at the two frequencies for the fixed $\beta =
10^{-3}$. For the `Outside' case, the width of the sample is fixed
at $W=20$. The sample size is varied by adjusting the number of
the cylinders. A few important features are discovered.

For $ka = 0.01$, the transmission decays exponentially with the
sample size for both `Inside' and `Outside' situations. In the
`Outside' case, there are two decay slops, i.~e. -0.0427 and
-0.0019. There is a transition regime separating the two slops.
The transmission starts to decay with slop $-0.0427$, followed by
a milder decay of slop -0.0019. The slop of $-0.0427$ is nearly
the same as the decay slope of $-0.0460$ in the `Inside' case.
Obviously, this is because at $ka=0.01$, waves are localized. This
exponential decay actually indicates that waves are trapped or
localized near the transmitting source; this is clearly shown by
the top portion of Fig.~\ref{fig3}(b). The second slop in the
`Outside' case for $ka=0.01$ is due to the finite width of the
sample, as will be discussed later. Inside the localization
regime, the transmission fluctuation is small, as expected from an
earlier work \cite{EY}. Here we see that within the localization
regime, wave localization can be observed in both `Outside' and
`Inside' scenarios.

\input epsf.tex

For $ka=0.05$, the `Inside' and `Outside' scenarios differ
significantly. While for the `Outside' case the transmission
decreases exponentially with a slop of 0.0027 along the path, the
transmission in the `Inside' situation does not decrease. For the
`Outside' case, the transmission fluctuation increases then drops
along the path, emulating the localization effect. For the
`Inside' case, however, the transmission fluctuation representing
the diffusive intensity\cite{Ishimaru} increases as more and more
scattering occurs along the path, fully complying with the
well-known non-localized Milne diffusion. The large fluctuation
implies that the transmission is sensitive to the distribution of
the cylinders, another indication of the non-localization
property\cite{EY}. In fact, the apparent decay in the `Outside'
case is due to the scattering attenuation that the waves are
reflected and scattered to the sides. These results suggest that
waves are actually not localized for $ka=0.05$, and it would be a
mistake to interpret the exponential decay shown in the `Outside'
situation as the indication of wave localization.

\begin{figure}
\epsfxsize=3in\epsffile{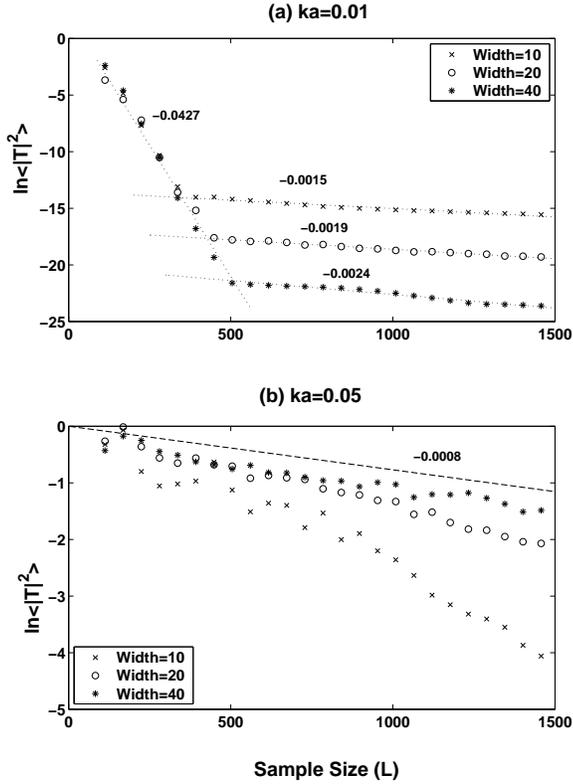} \caption{Normalized acoustic
transmission across a scattering array as a function of the sample
length for different sample widths. The solid line in (b) is the
numerical extrapolation as the width approaches infinity.}
\label{fig4}
\end{figure}

Now we consider the width effect in the `Outside' situation. In
Fig.~\ref{fig4}, the ensemble averaged transmission is plotted as
a function of the sample size (L) for three widths (W) at two
frequencies.

\input epsf.tex

For frequencies inside the localization regime, the second slop is
due to the finite width, and can be interpreted as follows. At
$ka=0.01$, the localization effect is dominant, thus leading to a
rapid decay in transmission along the path, giving rise to the
first slop. As expected, this slop is almost independent of the
width. As the sample size increases, the directly transmitted
waves are almost completely blocked. But, due to the finite width,
a small amount of the deflected waves around the sample sides can
still reach the receiver. When increasing the sample size (L) for
a fixed width (W), this effect gradually diminishes along the
path, yielding the second slop. Increasing width (W) will reduce
this finite-width effect, and thus the amount of deflected waves
reaching the receiver will also decrease, as shown. The width
effect disappears when the width approaches infinity. For
$ka=0.05$, outside the localization regime, the transmission
decays nearly exponentially along the path. As the width
increases, the slop will become smaller, and will be saturated to
a value of -0.0008. These results indicate that in the `Outside
scenario', the path-dependent transmission behaves similarly for
frequencies either inside or outside the localization regime as
long as the width is sufficiently large; therefore the phenomenon
of wave localization cannot be isolated in this scenario.

Finally, we stress that whether waves are localized or extended is
an intrinsic property of the system that is supposed to be
infinite. This property does not depend on the source, and should
not depend on boundaries either. We believe that while the source
is placed inside the medium with increasing sizes, the infinite
system might be mimicked and the localization property could be
probed without ambiguity.

%The work received support from National Science Council of
%Republic of China.


\begin{references}

\bibitem{Ishimaru} A. Ishimaru, {\it Wave propagation and scattering in
random media}, (Academic Press, New York, 1978).

\bibitem{Laser} N. M. Lawandy, R. M. Balachandran, A. S. L. Gomes, and
E. Sauvain, Nature {\bf 368}, 436 (1994); H. Cao, {\it et al.},
Phys. Rev. Lett. {\bf 82}, 2278 (1999).

\bibitem{Im} P. W. Anderson, Phys. Rev. {\bf 109}, 1492 (1958).

\bibitem{Band}
{e.~g.} E. Yablonovitch, Phys. Rev. Lett. {\bf 58}, 2059 (1987);
S. John, Phys. Rev. Lett. {\bf 58}, 2486 (1987); W. Robertson, et
al., Phys. Rev. Lett. {\bf 68}, 2023 (1992).

\bibitem{Sanchez} J. V. S\'anchez-P\'erez, et al., Phys. Rev.
Lett. {\bf 80}, 5325 (1998).

\bibitem{Kush} M. S. Kushwaha, Int. J. Mod. Phys. {\bf B10}, 977
(1996).

\bibitem{Kirk1} T. R. Kirkpatrick, Phys. Rev. {\bf B31}, 5746 (1985).

%\bibitem{Condat} C. A. Condat, J. Acoust. Soc. Am. 83, 441 (1988).

%\bibitem{Sor} D. Sornette and B. Souillard, Europhys. Letts. 7, 269
%(1988).

\bibitem{Genack} A. Z. Genack and N. Garcia, Phys. Rev. Lett. {\bf 66}, 2064
(1991).

\bibitem{Microwave} R. Dalichaouch, J. P. Amstrong, S. Schultz,
P. M. Platzman, and S. L. McCall, Nature {\bf 354}, 53 (1991); A.
Z. Genack and N. Garcia, Phys. Rev. Lett. {\bf 66}, 2064 (1991).

\bibitem{Ad} A. Lagendijk and B. A. van Tiggelen, Phys. Rep. {\bf 270}, 143
(1996).

%\bibitem{Sajeev} S. John, Physics Today, {\bf 44}(5), 32 (1991).

\bibitem{weak1} M. van Albada, and A. Lagendijk, Phys. Rev. Lett. {\bf 55}, 2692
(1985); P. E. Wolf, and G. Maret, Phys. Rev. Lett. {\bf 55}, 2696
(1985).

%\bibitem{weak2} A. Tourin, et
%al., Phys. Rev. Lett. 79, 3637 (1997).

%\bibitem{Sheng1} P. Sheng, {\it Introduction to wave scattering, localization
%and mesoscopic phenomena}, (Academic Press, New York, 1995).

\bibitem{Light} D. S. Wiersma, P. Bartolini, A. Lagendijk, and
R. Roghini, Nature {\bf 390}, 671 (1997).

\bibitem{Argue1} e. g. F. Scheffold, R. Lenke, R. Tweer, and G. Maret, Nature
{\bf 398}, 206 (1999); D. S. Wiersma, P. Bartolini, A. Lagendijk,
and R. Roghini, Nature {\bf 398}, 207 (1999).

\bibitem{Argue2} A. A. Chabanov, M. Stytchev, and A. Z. Genack,
Nature, {\bf 404}, 850 (2000).

%\bibitem{Phil} e.~g. P. W. Anderson, Phil. Mag. {\bf B 52}, 505 (1985).

\bibitem{gang4} E. Abrahams, P. W. Anderson, D. C. Licciardello, and
T. V. Ramakrishnan, Phys. Rev. Lett. {\bf 42}, 673 (1979).

\bibitem{EY} E. Hoskinson, and Z. Ye, Phys. Rev. Lett.
       {\bf 83}, 2734 (1999)

\bibitem{Ye3} Z. Ye, and E. Hoskinson, Appl. Phys. Lett. {\bf 77},
4428 (2000).

\bibitem{Twersky} V. Twersky, J. Acoust. Soc. Am. {\bf 24}, 42
(1951).

\bibitem{Yep} Y.-Y. Chen, and Z. Ye, Phys. Rev. {\bf E64}, 036616 (2001).

\bibitem{addition} I. S Gradshteyn, I. M. Ryzhik,
and A. Jeffrey, {\it Table of Integrals, Series, and Products},
5th Ed., (Academic Press, New York, 1994).

\end{references}
\end{document}